\documentclass[aps,pra,twocolumn,amsmath,amssymb]{revtex4-1}
\usepackage[colorlinks,pdfusetitle,urlcolor=blue,citecolor=blue,linkcolor=blue,bookmarksnumbered,plainpages=false]{hyperref}
\usepackage{graphicx}
\usepackage{courier}
\newcommand{\B}{\mathbf}
\usepackage{xcolor}

\newcommand{\tens}{\B}

\renewcommand{\i}{i}
\newcommand{\iG}{\text{Im}G(\B{r},\B{r},\omega)}
\newcommand{\frob}{\odot}
\newcommand{\dmu}{\B{K}}
\newcommand{\ddmumu}{\B{N}}
\usepackage{bm}
\usepackage{braket}
\usepackage[percent]{overpic}
\usepackage{stackengine}

\newcommand{\sep}{R}

\newcommand{\note}[1]{}
\begin{document}

\title{Inverse design of environment-induced coherence}
\author{Robert Bennett}
\affiliation{School of Physics \& Astronomy, University of Glasgow, Glasgow G12 8QQ, United Kingdom}
\date{\today}

\begin{abstract} 
Atomic transitions with orthogonal dipole moments can be made to interfere with each other by the use of an anisotropic environment. Here we describe, provide and apply a computational toolbox capable of algorithmically designing three-dimensional photonic environments that enhance the degree of coherence in atomic $\Lambda$ systems. Example optimisation runs yield approximately double the degree of coherence found using simple planar geometries. 
\end{abstract}

\maketitle

The interplay of transitions to and from sets of degenerate energy levels is responsible for a wide variety of well-established physical processes including lasing without inversion \cite{Scully1989}, populating trapping \cite{Arimondo1976}, quantum beats \cite{Dodd1964} and narrowing of spectral lines \cite{Zhou1996}. In order for two transitions to exhibit mutual coherence in the absence of external influences, they must have dipole moments that are non-orthogonal. This can be engineered in some specific situations \cite{Dutt2005}, but dipole moments for degenerate transitions within one quantum system do not typically satisfy this criterion \cite{Ficek2002}. However, almost two decades ago it was established that an anisotropic environment can induce coherence between transitions whose orthogonal dipole moments would otherwise forbid this \cite{Agarwal2000}. Building on the simple example of parallel plates discussed in Ref.~\cite{Agarwal2000}, a variety of works have sought to design environments that maximise this effect (see, for example, Refs \cite{Li2001,Jha2015b,Hughes2017,Lassalle2019,Hughes2017a,Jha2018}). 

One approach whose potential for optimising coherence has not yet been explored is inverse design. This is a recent direction in nanophotonics \cite{Jensen2011,Molesky2018} where dielectric structures are algorithmically designed in such a way that a given observable is extremised. The resulting structures have been experimentally proven to offer much greater performance than their `by hand' counterparts \cite{Su2018}. Recently, a formulation of inverse design particularly suited to dealing with light-matter interactions was put forward \cite{Bennett2019c}. Environment-induced coherence is, at its core, a light-matter interaction meaning the approach presented in Ref.~\cite{Bennett2019c} is immediately applicable. Inverse design as a general strategy is particularly suited to optimising environment-induced coherence since it is a process that relies on enhancing correlations between two transitions while simultaneously suppressing their individual spontaneous decay rates. These competing requirements mean that it is not at all clear how best to design a structure to do this for a given set of physical and engineering constraints. Allowing it to be done algorithmically is therefore a natural avenue to pursue. 

This article is structured as follows. In section \ref{EnvIndCoherenceIntro} we briefly summarise the basic expressions for coherence induced by an anisotropic quantum vacuum, and evaluate them for a simple planar geometry. In section \ref{InverseDesignSection} we move on to inverse design, beginning in \ref{CoherenceOptTheorySection} with a derivation of the gradient of the objective function we require. In section \ref{ImplementationSection} we provide details of the computational implementation and present some example results, followed by a discussion and comparison with previous work in section \ref{DiscussionSection}. Conclusions and directions for future work are given in section \ref{Conclusions}.

\section{Coherence and the anisotropic vacuum}\label{EnvIndCoherenceIntro}

\begin{figure}
\includegraphics[width = 0.6\columnwidth]{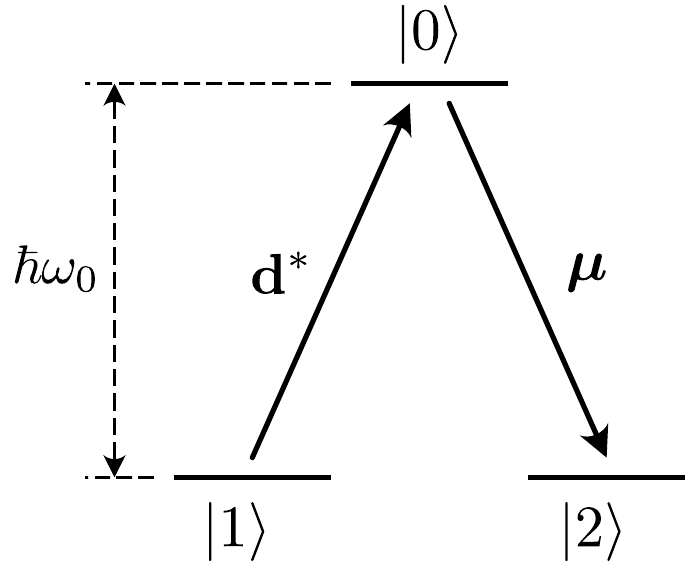}
\caption{Level scheme of the $\Lambda$ system considered here. \note{LambdaSystem.afdesign}  }
\label{Lambda}
\end{figure}

Consider a three-level quantum emitter with a $\Lambda$ structure as shown in Fig.~\ref{Lambda}.
Two nearly-degenerate ground states $\ket{1}$ and $\ket{2}$ are connected by transition dipole moments $\B{d}$ and $\bm{\mu}$ to an upper state $\ket{0}$, with energy splitting $\hbar \omega_0$. This type of system is physically realised in, for example, hyperfine levels of cold atoms. The master equation for the time-evolution of the atom's density matrix $\rho(t)$ can be written in the basis of its energy eigenstates as \cite{Lassalle2019};
\begin{align}\label{MasterEq}
\dot{\rho}(t) = - &  \left[ i\omega_0 + \frac{\gamma_1}{2} + \frac{\gamma_2}{2} \right] \ket{0} \bra{0} \rho(t) \notag \\
&\quad+ \rho_{00}(t) \Big[ \frac{\gamma_1}{2} \ket{1} \bra{1} + \frac{\gamma_2}{2} \ket{2} \bra{2} \notag \\
&\qquad+ \frac{\kappa_{21}}{2} \ket{2} \bra{1} + \frac{\kappa_{12}}{2} \ket{1} \bra{2}\Big] + \text{H.c.,}
\end{align}
where $\rho_{00}(t)$ is the population of the upper state, $\gamma_1$ and $\gamma_2$ are respectively the spontaneous decay rates from the upper state to states $1$ and $2$;
\begin{align}\label{gammas}
\gamma_1 &= \frac{2 \omega_0^2}{\hbar \varepsilon_0 c^2} \B{d}^* \cdot \text{Im} \B{G}(\B{r},\B{r},\omega_0) \cdot \B{d},\notag \\
\gamma_2 &= \frac{2 \omega_0^2}{\hbar \varepsilon_0 c^2} \bm{\mu}^* \cdot \text{Im} \B{G}(\B{r},\B{r},\omega_0) \cdot \bm{\mu},
\end{align}
and $\kappa_{12}$ is the coupling between the two degenerate transitions
\begin{equation}\label{kappa}
\kappa_{12} = \frac{2 \omega_0^2}{\hbar \varepsilon_0 c^2} \B{d}^* \cdot \text{Im} \B{G}(\B{r},\B{r},\omega_0) \cdot \bm{\mu}. 
\end{equation}
In these expressions $\tens{G}(\B{r},\B{r}',\omega)$ is the dyadic Green's tensor describing propagation of polaritons (or photons when in free space) from position $\B{r}'$ to  $\B{r}$ at angular frequency $\omega$. This tensor depends on the geometry and materials of the environment, which, as we shall see, need to be different from vacuum in order to induce coherence.

The steady-state values of the off-diagonal elements of the density matrix whose time evolution is governed by Eq.~\eqref{MasterEq} are \cite{Agarwal2000,Lassalle2019};
\begin{equation} \label{rho12BasicDef}
\rho_{12}(t\to\infty) = \rho_{21}^*(t\to\infty) = \frac{\kappa_{12}}{\gamma_1 + \gamma_2} \equiv \rho_{12} ,
\end{equation}
the absolute value of which we will seek to maximise. It is helpful for later calculation to convert \eqref{rho12BasicDef} into the following form;
\begin{equation}\label{rho12NiceForm}
\rho_{12} = \frac{ \dmu \odot \text{Im} \B{G}(\B{r},\B{r},\omega_0)}{\ddmumu \odot \text{Im} \B{G}(\B{r},\B{r},\omega_0)} 
\end{equation}
where
\begin{align}\label{MatrixDefinitions}
 \dmu &\equiv \B{d}^* \otimes \bm{\mu}, & \ddmumu &\equiv \B{d}^* \otimes \B{d} + \bm{\mu}^* \otimes \bm{\mu}.
\end{align}
The trace of the matrix $\dmu$ is equal to the inner product of the dipole moments;
\begin{equation}\label{TrK}
\text{Tr} \dmu = \B{d}^* \cdot \bm{\mu}
\end{equation}
so is simply a measure of the orthogonality of the pair of transitions. 

\subsection{Vacuum}\label{VacuumSection}

In vacuum the imaginary part of the equal-point Green's tensor is proportional to a unit matrix [see Eq.~\eqref{ImGVac}] under which conditions the coherence becomes;
\begin{equation} \label{rho12Vac}
\rho_{12} = \frac{\text{Tr} \B{K}}{\text{Tr}\B{N}} = 0
\end{equation}
with the second equality holding via Eq.~\eqref{TrK} if the dipole moments are orthogonal. This is a demonstration of the well-known fact that orthogonal dipole transitions are uncorrelated in vacuum (see, for example, \cite{Agarwal2000}). 

\subsection{Perfect reflector}

The Green's tensor is no longer proportional to an identity matrix if an anisotropic environment is introduced, so Eq.~\eqref{rho12Vac} no longer holds in this case. The simplest example of an inhomogeneous environment  is a perfectly reflecting plane positioned in, say, the $xy$ plane, for which the imaginary part of the equal-point Green's tensor ($\mathbf{r} = \mathbf{r}'$) on the $z$-axis is (see Appendix \ref{GreensTensorAppendix});
\begin{align}\label{ImGPM}
&\text{Im} \tens{G}(\B{r},\B{r},\omega)=\frac{\omega}{6\pi c}\mathbb{I}_3\notag\\
&+ \frac{\left(1-4 \pi ^2 \zeta_z ^2\right) \sin (2 \pi  \zeta_z )-2 \pi  \zeta_z  \cos (2 \pi  \zeta_z )}{32 \pi ^3 \zeta_z ^2 z}\text{diag}(1,1,0)\notag  \\
&+ \frac{\sin (2 \pi  \zeta_z )-2 \pi  \zeta_z  \cos (2 \pi  \zeta_z )}{16 \pi ^3 \zeta_z ^2 z} \text{diag}(0,0,1)
 \end{align}
where $\zeta_z = \omega z/{\pi c}$ is a dimensionless parameter, the choice of which will be motivated at the end of this section. The translational symmetry of this environment in the $xy$-plane is reflected in the Green's tensor by\eqref{ImGPM} being diagonal in its upper left block, so choosing the dipole moments to rotate in the $xy$ plane results in vanishing coherence, just like in vacuum. This behaviour has a clear physical interpretation, since the downward dipole transition $\bm{\mu}$ emits light of (say) left-circular polarisation which is converted to right-circular polarisation upon reflection by the interface (as viewed along its own optical axis), but remains left-circular from the perspective of the atom. This means it cannot excite the right-circular transition $\B{d}$.

\begin{figure}
\centering
\includegraphics[width=\columnwidth]{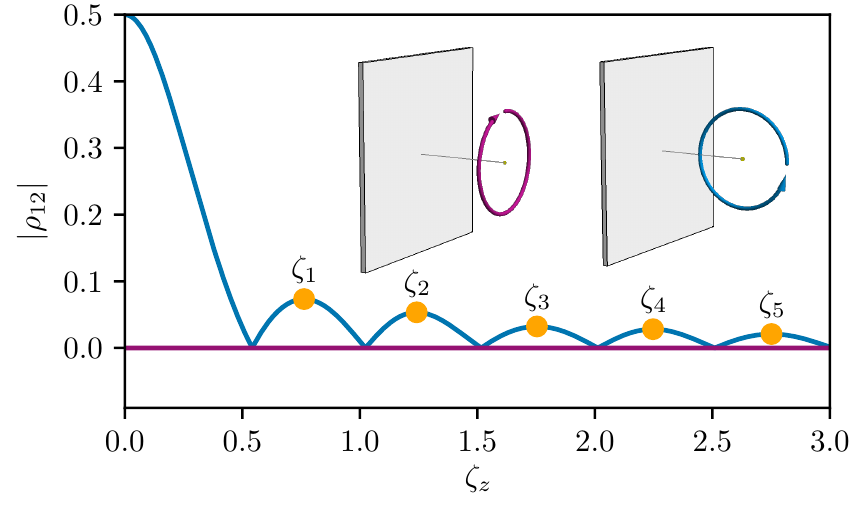}
  \caption{\note{HalfSpaceComparison.py} Absolute value of the atomic coherence at a distance $\zeta_z=\omega z/\pi c$ from a perfectly reflecting surface. For in-plane rotation (purple), the coherence vanishes, while for perpendicular rotation the coherence displays characteristic oscillations. The position of the first antinode (aside from the one at the surface itself) indicated by the orange dot is at $\zeta_z \approx 0.7627$. This is the point at which all the fixed-$\zeta_z$ simulations in the rest of this work are undertaken. }
\label{PMPlot}
\end{figure}

In order for $\rho_{12}$ to be non-zero we therefore need $\B{d}$ and $\bm{\mu}$ to have non-zero components in the $z$ direction, as well as in either the $x$ or $y$ direction. For this example we choose the latter, taking the orthogonal dipole moments as;
\begin{align}\label{yzDipoles}
\B{d} &= \frac{d}{\sqrt{2}} \{ 0,1, \i  \} & \bm{\mu} &=\frac{\mu}{\sqrt{2}} \{0,1, -\i\}
\end{align}
where $d$ and $\mu$ are real constants. The matrices and $\dmu$ and $\ddmumu$ then follow directly from their definitions \eqref{MatrixDefinitions}, plugging these together with the Green's tensor \eqref{ImGPM} into Eq.~\eqref{rho12NiceForm} one finds for the coherence induced by the perfect reflector;
\begin{align}\label{PMResult}
&\rho_{12} =\frac{2d \mu} {|\mu|^2+|d|^2}\notag \\
&\times \frac{6 \pi  \zeta_z  \cos (2 \pi  \zeta_z )-3 \left(4 \pi ^2 \zeta_z ^2+1\right) \sin (2 \pi  \zeta_z )}{(4\pi \zeta_z) ^3-6 \left(4 \pi ^2 \zeta_z ^2-3\right) \sin (2 \pi  \zeta_z )-36 \pi  \zeta_z  \cos (2 \pi  \zeta_z )}\end{align}
The absolute value of this for the case $d=\mu$ is plotted in Fig.~\ref{PMPlot},
where it is in general different from zero. This again has a clear physical interpretation --- the light emitted towards a mirror by a dipole rotating perpendicularly to it is \emph{linearly} polarised. Thus, provided it has an appropriate phase after reflection, it can be absorbed by a dipole rotating in the opposite direction. This phase requirement is demonstrated in Fig.~\ref{PMPlot} by the fact that the coherence oscillates with a period determined by $\zeta_z=\omega z/\pi c = 2z/\lambda$. This dimensionless quantity represents the round-trip distance to the surface in units of the wavelength $\lambda$ --- when the emitter is at any position satisfying $\zeta_z = n/2$ for an integer $n$, the coherence vanishes as should be expected from destructive interference.  The exact positions of the peaks in the coherence $\zeta_n$ are given by the solution of a transcendental equation, but obey  $\zeta_n \approx \frac{1}{2}(n+\frac{1}{2})$ to an accuracy of around 2\% (e.g. $ \zeta_1 = 0.7627\ldots$). The use of the dimensionless variable $\zeta_z$ emphasises that the results presented here and in subsequent sections do not depend on particular choices of frequency and distance scale, rather one implies the other for a given value of $\zeta_z$. For example, taking the caesium D2 frequency of $\omega = 2\pi \times 352$THz, then $\zeta_z = 1$ represents $z= c/(704 \text{THz}) \approx 426$nm

\section{Inverse design}\label{InverseDesignSection}

The perfect reflector in the previous section was chosen as it is the simplest example of an inhomogeneous environment capable of inducing coherence. However, there is of course no reason that it should be in any sense optimal, or even good. For example, Fig.~\ref{PMPlot} demonstrates that the coherence falls away quite rapidly after the initial at-surface maximum. The task is then to systematically determine a choice of input to the model (e.g. a particular geometry) that gives a (locally) optimal output, which is known in general as inverse design.

\begin{figure}
\includegraphics[width = 0.72\columnwidth]{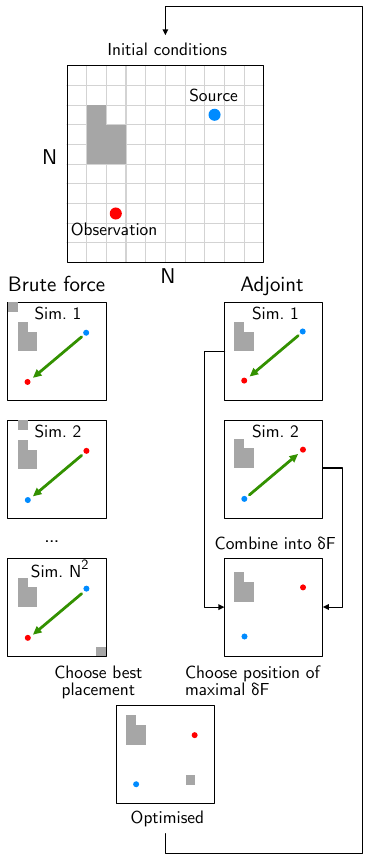}
\caption{\note{BlockPlacing.afdesign} Schematic illustration of the difference between brute force and adjoint optimisation techniques, and the overall iterative approach taken here. }
\label{BlockPlacing}
\end{figure}

A brute force search of the vast parameter space of possible environments is computationally infeasible, to avoid this we use iterative adjoint optimisation \cite{Jameson1988}, which exploits the source-observer symmetry of Maxwell's equations to massively reduce numerical overhead. As schematically illustrated in Fig.~\eqref{BlockPlacing}, 
it requires (at most) \emph{two} calculations of $\B{G}$ to find the position that a piece of material should be placed in order to maximally increase a given observable $f$, rather than having to simply repeat the calculation for placement at every possible point in the simulation volume and select the best result. The key quantity that tells us the optimal placement position $\B{r}''$ for an observable depending on the Green's tensor at two positions $\B{r}$ and $\B{r}'$ is the merit function gradient $\delta F$, given by \cite{Bennett2019c}: 
\begin{align} \label{BasicDF}
\delta F &= \text{Re} \left[ \frac{\partial f}{\partial \tens{G}}(\B{r},\B{r}',\omega)\frob \tens{G}^\mathrm{T}(\B{r}'',\B{r},\omega ) \cdot \tens{G}(\B{r}'',\B{r}',\omega )\right].
\end{align}
where $\B{G}$ and its conjugate are treated as independent variables. Positions $\B{r}$ and $\B{r}'$ are fixed, so Eq.~\eqref{BasicDF} can be fully determined by calculating $\tens{G}$ for all observation positions given a source placed at $\B{r}$ and again for a source placed at $\B{r}'$. This is the origin of the reduction to two simulations from $N^2$ (or $N^3$ in 3D) required in a brute force approach. Here and throughout we ignore all real, positive prefactors appearing in the merit function gradient $\delta F$ without further comment, as these make no difference to the spatial positions of its zeros or of its maximum, which are the only quantities we are interested in. 

The technique of adjoint optimisation brings the problem well within computational reach, so is the approach taken here. In the particular example of environment-induced coherence, the source and observation point happen to be the same, so in this case we need only do one simulation per iteration.

\subsection{Optimising coherence}\label{CoherenceOptTheorySection}

To tackle our particular problem of optimising $|\rho_{12}|$ given by Eq.~\eqref{rho12NiceForm} we simple choose $f = |\rho_{12}|$ in Eq.~\eqref{BasicDF}. Expression of $\delta F$ in terms of $\tens{G}$ then entails calculation of the following functional derivative 
\begin{equation}\label{functionalDrv}
\frac{\partial}{\partial \B{G}} |\rho_{12}|=\frac{1}{|\rho_{12}|} \text{Re}\left( \rho^*_{12}\frac{\partial \rho_{12}}{\partial \B{G}} \right).
\end{equation}
After some algebra, one finds;
\begin{equation}
\frac{\partial \rho_{12}}{\partial \B{G}} =\frac{1}{2i} \frac{\dmu(\ddmumu \odot \B{\iG}) - \ddmumu(\dmu\odot \B{\iG})}{[\ddmumu\odot \B{\iG}]^2}
\end{equation}
which can then be used in Eq.~\eqref{functionalDrv}, giving;
\begin{align} \label{MeritFnCoherence}
\delta F =&\text{Re}\Bigg\{\frac{1}{2i}\left|\frac{ \ddmumu \odot \text{Im} \B{G}(\B{r},\B{r},\omega)}{\dmu \odot \text{Im} \B{G}(\B{r},\B{r},\omega)}  \right|  \left[\frac{ \dmu \odot \text{Im} \B{G}(\B{r},\B{r},\omega)}{\ddmumu \odot \text{Im} \B{G}(\B{r},\B{r},\omega)}  \right]^*\notag \\
& \times\frac{\dmu[\ddmumu \odot \B{\iG}] - \ddmumu[\dmu\odot \B{\iG}]}{[\ddmumu\odot \B{\iG}]^2}\notag \\
& \frob \tens{G}^\mathrm{T}(\B{r}'',\B{r},\omega ) \cdot \tens{G}(\B{r}'',\B{r},\omega )\Bigg\}
\end{align}
This expression simplifies considerably when the vacuum Green's tensor \eqref{GVac} is used, becoming
\begin{equation}
\frac{\partial \rho_{12}}{\partial \B{G}} =\frac{1}{12\i \pi c} \frac{\dmu \text{Tr} \ddmumu-\ddmumu \text{Tr} \dmu}{(\text{Tr} \ddmumu)^2}=\frac{1}{12\i \pi c} \frac{\dmu }{\text{Tr} \ddmumu}
\end{equation}
where on the right hand side we used that $\text{Tr}\dmu = 0$ for orthogonal dipole moments [see Eq.~\eqref{TrK}]. Consequently, the merit function change in vacuum is:

\begin{align}\label{dFVac}
&\delta F_\text{vac} =  \text{Re}\left[\frac{\dmu }{i\text{Tr} \ddmumu} \frob \tens{G}^\mathrm{T}(\B{r}'',\B{r},\omega ) \cdot \tens{G}(\B{r}'',\B{r},\omega )\right]
\end{align}
where we have also used that $\text{Tr} \ddmumu$ is necessarily real and positive, see Eq.~\eqref{MatrixDefinitions}.

\begin{figure}
\includegraphics[width = 0.8\columnwidth]{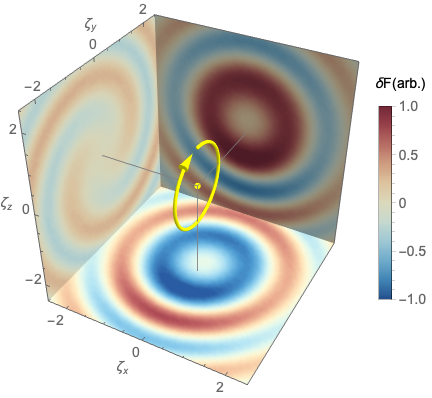}
\caption{\note{VacuumTest.nb} Spatial merit function for environment induced coherence for an atom at the origin with dipole moment rotating in the $yz$ plane, as indicated. Red regions indicate those where a piece of material would increase coherence, while blue regions are those which would suppress it. The value of the merit function is normalised to the largest (positive) value found across all of the three cross-sections shown. The structure in plane of rotation is strongly reminiscent of the spiral and `gammadion' structures found to exhibit highly chiral response \cite{Schnell2016,Paterson2018,Rodier2019}.}
\label{dF3DPlot}
\end{figure}

Equation~\eqref{dFVac} gives us our first insight into how we may go beyond planar surfaces in optimising coherence. To see this we place the atom at the origin and assume without loss of generality that the dipole moments are given by \eqref{yzDipoles}. The merit function gradient in this case becomes:
\begin{align} \label{dFVacExplicit}
\delta F_\text{vac} =&2 \Big[\left(\chi ^4+\chi ^2-3\right) \cos (2 \chi ) \notag \\
&\qquad \qquad -2 \chi  \left(\chi ^2+3\right) \sin (2 \chi )\Big] \zeta _y'' \zeta _z''\notag \\
&+\Big[2 \chi  \left(\chi ^2+3\right) \cos (2 \chi )\notag \\
&\qquad +\left(\chi ^4+\chi ^2-3\right) \sin (2 \chi )\Big] \left(\zeta _z''^2-\zeta _y''^2\right)
\end{align}
where we have introduced 
\begin{align}
\chi &= \pi \sqrt{\zeta_x''^2 + \zeta _y''^2 + \zeta_z''^2}\\
\{\zeta_x'',\zeta_y'',\zeta_z''\} &= \frac{\omega}{\pi c} \{x'',y'',z''\}
\end{align}
A plot of $\delta F$ as a function of $\zeta_x'',\zeta_y''$ and $\zeta_z'' $ is shown in Fig.~\ref{dF3DPlot}, 
from which we can draw several qualitative conclusions about the optimisations to be carried out. Firstly, structures in the plane of rotation have a spiral character, familiar from a class of chiral metasurfaces \cite{Schnell2016,Paterson2018,Rodier2019}.
Secondly, optimisation in the plane perpendicular to the plane of rotation is expected to be more effective than that parallel to it since the relative magnitude of $\delta F$ is much larger there. 

Placing a small block of dielectric material at the point of maximum $\delta F$ would increase $|\rho_{12}|$, but only very modestly. To find significant improvements, one has to take the environment as including this first block and determine the optimal placement of the \emph{next} block and so on --- the process becomes iterative. It is important to note that as soon as a piece of material is placed anywhere in the environment, it is of course no longer vacuum so a new Green's tensor must be calculated. This, in general, must be done numerically since the Green's function is only expressible analytically for planes, cylinders and spheres (as well as layered versions thereof, see for example \cite{Chew1995}). Therefore the result \eqref{dFVacExplicit} represents the first and only analytic step in a procedure that must continue numerically.

In order to carry out the numerics, we use the free finite difference time domain (FDTD) package Meep \cite{Oskooi2010} to calculate the Green's tensors using the method discussed in \cite{Bennett2019c}. Briefly, to calculate $\tens{G}(\B{r},\B{r}',\omega)$ a point current source $\B{j}$ is introduced at $\B{r}'$ and the resulting electric field at the observation point $\B{r}$ is calculated. Dividing the resulting vector by the source current component-wise and Fourier transforming, one is furnished with one row of the Green's tensor (corresponding to whichever direction the source current was chosen to be aligned). Carrying out the same process for the remaining two rows then gives all nine components of the FDTD Green's tensor for a particular $\B{r}$, then the whole process can be repeated for each point in the grid of observation points required for evaluation of \eqref{MeritFnCoherence}. We emphasise here that there is only one source point $\B{r}$, so the Green's tensor only has to be calculated once in a given geometry to find optimal placement of the next block, in contrast to brute force optimisation where each position would have to be tried. The numerical nature of this method means discretisation error and possible artefacts needs to be accounted for and controlled, our methods for doing this are discussed in Appendix \ref{NumericalErrorSection}. 

\subsection{Implementation}\label{ImplementationSection}

In order to make the predicted structures more realistically manufacturable, we include an optional background geometry of a perfectly reflecting plane (referred to as the backplate), upon which the algorithm is allowed to place a layer of material. When no backplate is present the algorithm is subject to the same constraints, so it builds a free-standing planar structure. Four physical situations were then considered --- with/without the backplate and parallel/perpendicular rotation of the dipole moments, relative to the plane of optimisation. For parallel rotation the dipole moments are
\begin{align}
\B{d} &= \frac{d}{\sqrt{2}} \{ 1,\i,0\}  & \bm{\mu} &=\frac{\mu}{\sqrt{2}} \{1,-\i,0 \}
\end{align}
while for perpendicular rotation the dipole moments given by Eq.~\eqref{yzDipoles}. In all cases $d=\mu$ was assumed for simplicity, the coherence for for $d\neq \mu$ can be obtained from the values presented here by inserting a factor with ${2d \mu}/(|\mu|^2+|d|^2)$ [see Eq.~\eqref{PMResult}]. 

The physical parameters were chosen as follows. The material being placed by the algorithm at each step is a cube of side length $\lambda/6$ with permittivity $\varepsilon = 3$ (referred to as a block from here on) --- roughly corresponding to materials like glass or sapphire. The perfectly reflecting backplate has the same dimensions as the optimisation region, and is half a wavelength deep (although this is immaterial since by definition its thickness does not matter).  In the simulations with the backplate the atom was at the first antinode $\zeta_1$ measured relative to the vacuum/backplate interface (see Fig.~\ref{PMPlot}), and in the freestanding simulations it is the same distance but measured from the centre of the structure in the $\zeta_z$ direction. 

The computational parameters chosen were a resolution twelve pixels per wavelength, as this was found to result in a good tradeoff between accuracy and speed (see appendix \ref{NumericalErrorSection}). In each case the atom was placed on the $\zeta_z$ axis, the optimisation region was three wavelengths square in the $\zeta_x-\zeta_y$ plane and one block deep in $\zeta_z$, centered at the origin. The overall simulation box size is four wavelengths, and beyond this a set of perfectly matched layers ensure near-perfect absorption of any outgoing radiation. The computational parameters were confirmed as being sufficient by comparing with the analytic perfect reflector result \eqref{PMResult}, see appendix \ref{NumericalErrorSection}.

\begin{figure*}
 \includegraphics[width = \textwidth]{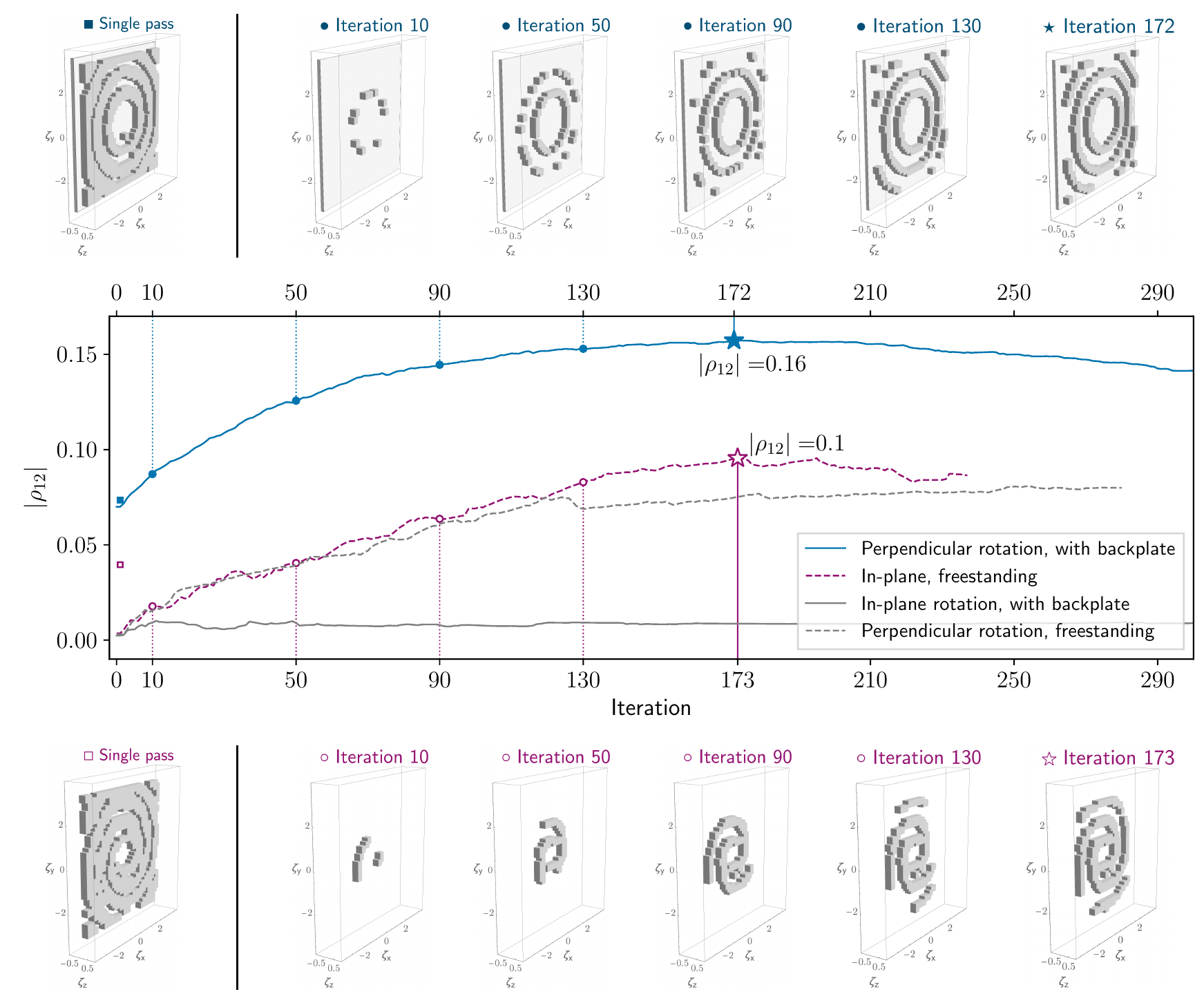}
\caption{\note{IterationProcessPlots.py, GeomVisualiserV2.nb, 12ppw} Main plot: absolute value of the atomic coherence at the first antinode from Fig.~\ref{PMPlot} ($\zeta_1 \approx 0.7627$) at each step of the iterative process, for four different physical situations (with/without backplate, parallel/perpendicular rotations), with parameters detailed in the main text.  Shown above and below the main plot are the structures at selected stages in two of the four optimisation runs (corresponding coherence marked by circles on the main plot), as well as the `single pass' structure found for each (marked by squares). The iterations at which the coherence peaked for those two runs are marked by stars, with the final structures shown in the upper and lower right. Each step in the optimisation took approximately ten minutes per core on the University of Glasgow High Performance Computing facility. }
\label{iterationProgression}
\end{figure*}

As a test of the necessity of the computationally-heavy process of iterative inverse design, we also investigated the coherence for what we term `single pass' design. This proceeds by beginning from vacuum, taking the analytic merit function as shown in Fig.~\eqref{dF3DPlot} and simply placing material at any position where $\delta F>0$. The coherence $\rho_{12}$ can then be evaluated with a single simulation. The results of the four iterative optimisation runs described in this section (as well as two single-pass results) are shown in Fig.~\ref{iterationProgression}. The code underpinning the simulations can be found at Ref.~\cite{Bennett2020data}, alongside detailed documentation. 

\subsection{Discussion}\label{DiscussionSection}

\begin{figure}
\includegraphics[width = \columnwidth]{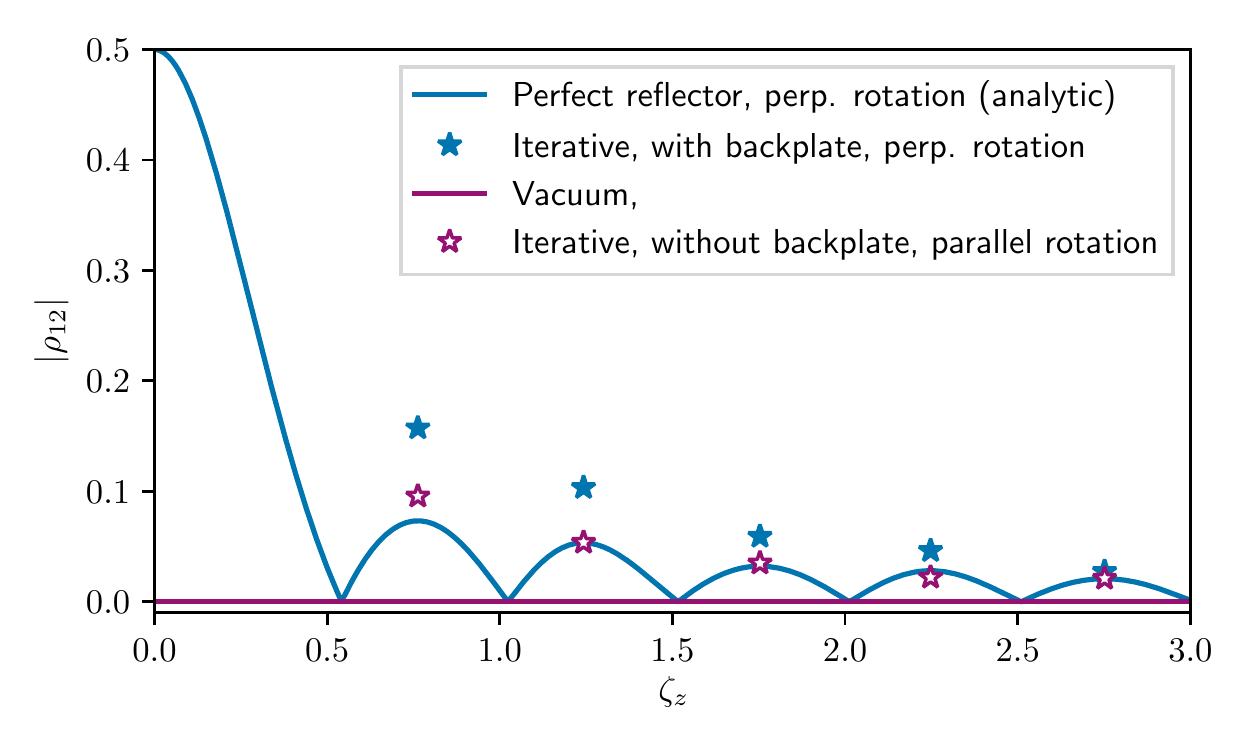}
\caption{\note{NodeComparison.py} Coherence induced by a perfect reflector and by vacuum (solid lines) and the results of iterative optimisation at the five anti-nodes. For the case of optimisation without the backplate (i.e. beginning in vacuum), 'parallel rotation' is meant as with respect to the plane in which the algorithm is allowed to place material.  }
\label{NodeComparison}
\end{figure}

The highest absolute coherence is found, perhaps unsurprisingly, by using the iterative optimisation technique for the case of perpendicular rotation with the backplate. This is because the starting structure already induces coherence in a similar way to the infinitely extended perfectly reflecting plane as shown in Fig.~\ref{PMPlot}. The inverse design algorithm patterns the surface in such a way to make this reasonably realistic compact structure induce approximately twice the degree of coherence as its infinitely extended (unphysical) counterpart. This conclusion holds in at points other than the first antinode $\zeta_1$ chosen in Fig.~\ref{iterationProgression} --- in Fig.~\ref{NodeComparison} we summarise the results of repeating the two simulations highlighted in Fig.~\ref{iterationProgression} for the remaining antinodes. 

The single-pass approach does not work as well as the iterative approach. This is because it is inconsistent with the assumptions under which the merit function gradient \eqref{BasicDF} was derived (addition of pieces of dielectric of with small optical volume), so there is no compelling reason the resulting structure should improve coherence (and could even reduce it). Nevertheless, it is significantly computationally cheaper, with only one numerical simulation required as opposed to hundreds. Its greatest success is found in free-standing optimisation for a dipole rotating in the parallel direction, since essentially any new material in the optimisation plane will break the translational symmetry that leads to vanishing coherence. By contrast, for a dipole rotating in the perpendicular direction with a backplate already present, the symmetry is already broken so further optimisation is more delicate. In both cases, the iterative method outperforms single pass optimisation, though much less dramatically in simulations with the backplate. 

We now briefly compare the results of this work to those of other enhancement techniques. In \cite{Lassalle2019}, a 1D resonant metasurface was designed and the coherence was reported as reaching approximately $0.1$ at a distance $20\zeta_z$ --- far in excess of the results for coherence presented here.  However, the authors of \cite{Lassalle2019} caution that they `do not take into account all the details of the metasurface', instead taking it as optically equivalent to an idealised spherical mirror modified by some transmissivity values for a single polarisation. The authors also assume the part of the decay rate that stems from the component of the dipole moment perpendicular to their metasurface's periodicity is unchanged. These assumptions may artificially enhance the coherence, whereas the full numerical treatment presented here \cite{Lassalle2019} is expected to be more realistic. The authors of \cite{Li2001} considered the situation of perpendicular rotation in a multilayer dielectric medium. For the case of an atom placed in vacuum between two dielectric slabs, they find values for the absolute coherence up to approximately 0.05 for atom-surface distances exceeding $\zeta_z \approx 2$. This is similar to the perfect reflector results presented here due to approximate cancellation of two competing effects (enhancement due to modes trapped between the slabs, suppression due to a lower reflectivity surface), and is thereby outperformed by the iterative techniques used here. Consideration of multilayer geometries like those in \cite{Li2001} will form the basis of future work. 

\section{Conclusions}\label{Conclusions}

In this work we have presented and applied a toolbox for using inverse design to optimise environment-induced coherence. We derived a very general merit function in terms of dyadic Green's tensors, and applied this to the case of vacuum in order to provide insight into what type of structures should induce coherence. We then used iterative inverse design to show that this method can enhance existing coherence by a factor of approximately two via simple surface patterning, as well as induce appreciable coherence in situations where there was none initially present. While the values found for the coherence do not exceed some previous claims for metasurfaces, the approach presented here is much more flexible that those preceding it. For example, neither the starting geometry nor the optimisation region are limited to being planar, either could be of any three-dimensional shape (e.g. spheres, gratings, parabolas). These will form directions for future work using the numerical tools developed here, available at \cite{Bennett2020data}. 

\acknowledgments{The author thanks Stephen Barnett and James Cresser for helpful comments on the manuscript, as well as the Glasgow High Performance Compute Cluster support team for technical assistance. } 

\appendix

\section{Green's tensors}\label{GreensTensorAppendix}

The Green's tensor $\tens{G}^{(0)}(\B{r},\B{r}',\omega) $ for free space is (see, for example, \cite{Buhmann2012BothBooks})
\begin{align}\label{GVac}
\tens{G}^{(0)}(\B{r},\B{r}',\omega) =& -\frac{1}{3k^2} \mathbb{I}_3 {\delta}^{(3)}(\B{\sep}) \notag \\
&\,\,\,- \frac{e^{ikR}}{4\pi k^2 \sep^3} \Big\{[1-ik\sep - (k\sep)^2]\mathbb{I}_3 \notag \\
&\quad  - [3 - 3 i k \sep -  (k\sep)^2] \hat{\B{\sep}} \otimes \hat{\B{\sep}} \Big\}
\end{align}
 where $k = \omega/c$,  $\B{\sep} = \B{r}-\B{r}'$ and $\sep = |\B{\sep}|$. The delta function in the first term causes this to be ill-defined at $\B{R}=0$, but its imaginary part remains finite and is given by;
\begin{equation}\label{ImGVac}
\text{Im}\tens{G}^{(0)}(\B{r},\B{r}',\omega) = \frac{\omega}{6\pi c} \mathbb{I}_3. 
\end{equation}

The Green's tensor for a planar surface of permittivity $\varepsilon$ and unit permeability in the plane $z=0$ is given for $z, z' >0$  by;
\begin{equation}
\tens{G}(\B{r},\B{r}',\omega) = \tens{G}^{(0)}(\B{r},\B{r}',\omega) + \tens{G}^{(1)}(\B{r},\B{r}',\omega)
\end{equation}
where
\begin{align}\label{G1}
 \tens{G}^{(1)}(\B{r},\B{r}',\omega) =& \frac{i}{8\pi^2} \sum_{\sigma = \text{s}, \text{p}} \int d^2 k_\parallel  \frac{1}{k_z} e^{i \B{k}_\parallel \cdot (\B{r}-\B{r}')}  \notag \\
& \times e^{i k_z (z+z')} r_\sigma \B{e}_{\sigma+} \otimes \B{e}_{\sigma-}
\end{align}
where 
\begin{align}
\B{e}_{s\pm} &= \hat{\B{k}}_\parallel \times \hat{z} & \B{e}_{p\pm} &= \frac{1}{k} (k_\parallel \hat{z} \mp \hat{\B{k}}_\parallel )
\end{align}
with $\B{k}_\parallel = \{k_x, k_y, 0 \}$, $k_\parallel = |\B{k}_\parallel|$ and, $r_s$ and $r_p$ being the Fresnel reflection coefficients for $s$ and $p$ polarisations. In general these coefficients depend on the wavevector $\B{k}$, but for a perfect reflector they are simply given by $r_s=-1$ and $r_p=1$. Substituting these values into \eqref{G1} and taking equal position arguments $\B{r} = \B{r}'$ allows the frequency integrals can be carried out. All off-diagonal elements vanish, and the diagonal elements are given by:
\begin{align}
G^{(1)}_{xx} = G^{(1)}_{yy} &= \frac{e^{2 i \pi  \zeta } \left(1-2 i \pi  \zeta-4 \pi ^2 \zeta ^2 \right)}{32 \pi ^3 \zeta ^2 z} \label{Gxx}\\
G^{(1)}_{zz} &= \frac{e^{2 i \pi  \zeta } (1-2 i \pi  \zeta )}{16 \pi ^3 \zeta ^2 z} \label{Gzz}
\end{align}
where we have again used the dimensionless parameter $\zeta_z = \omega z/{\pi c}$ introduced in the main text. Taking the imaginary part of the diagonal matrix defined by \eqref{Gxx} and \eqref{Gzz},  then adding the result to Eq.~\eqref{ImGVac} results in Eq.~\eqref{ImGPM} in the main text. 

\section{Convergence and validation}\label{NumericalErrorSection}

\begin{figure}
\includegraphics[width = \columnwidth]{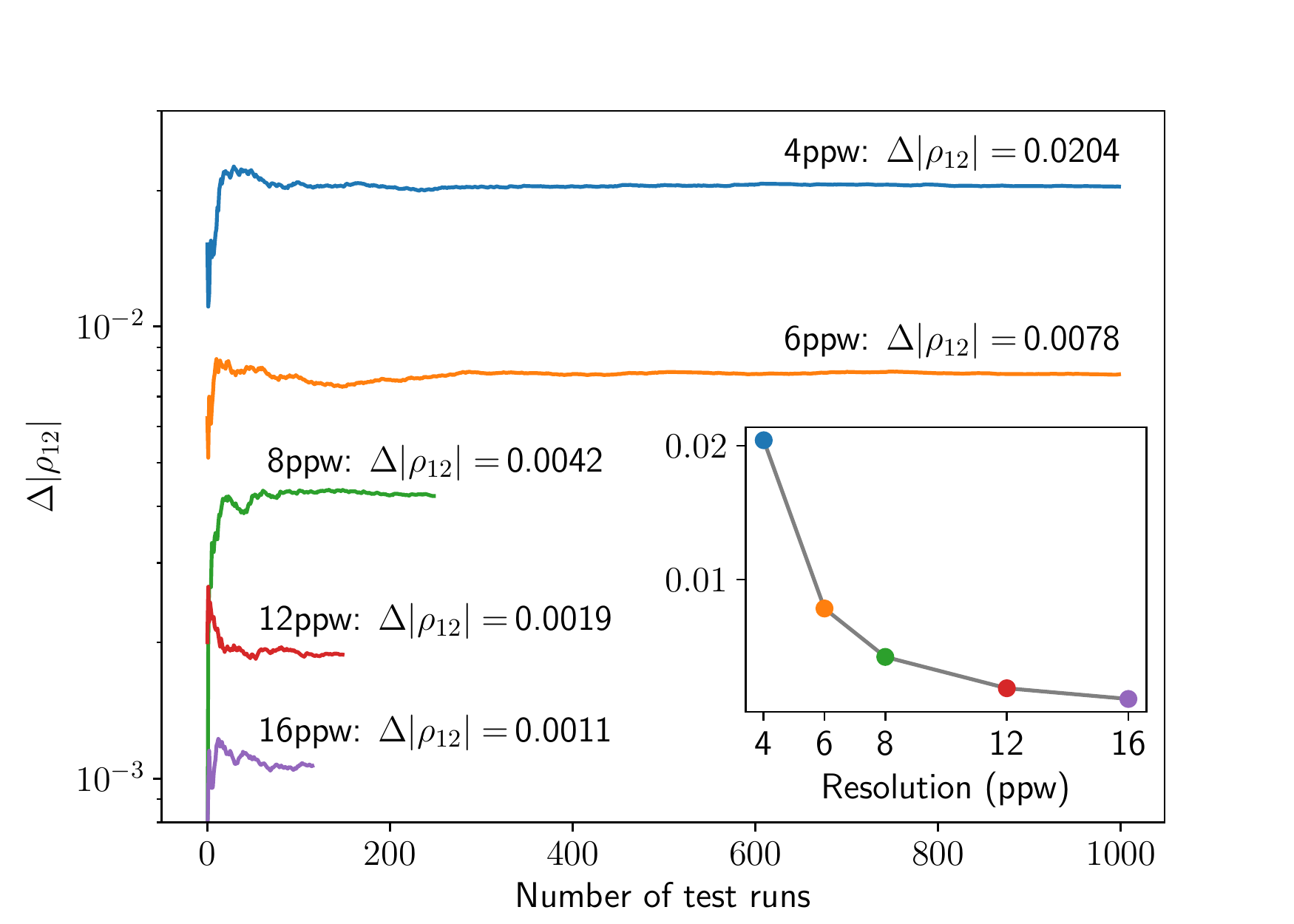}
\caption{\note{AccuracyDataAnalysis.py}  }
\label{resolutionPlot}
\end{figure}

\begin{figure}
\includegraphics[width = \columnwidth]{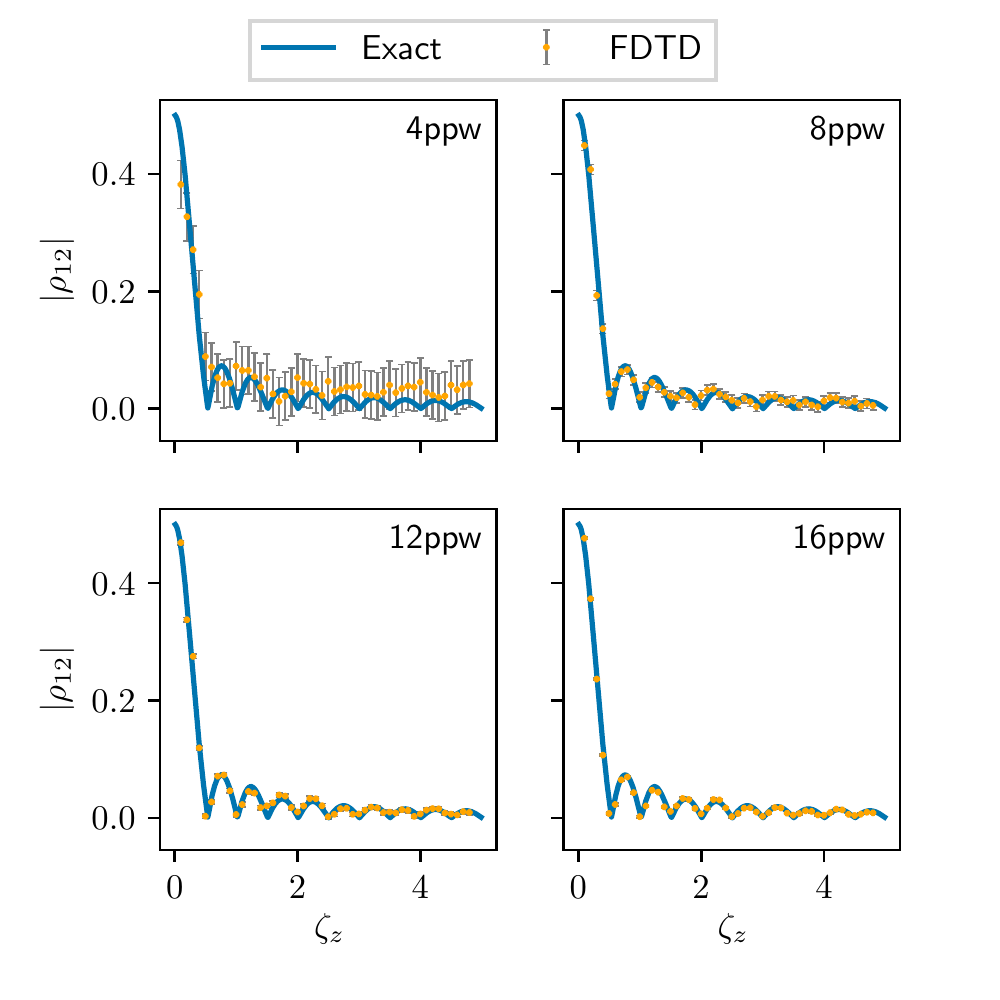}
\caption{\note{HalfspaceComparison.py} Exact results for the coherence induced by a perfectly reflecting half-space for perpendicular rotation [Eq.~\eqref{PMResult}] and the same calculated at various resolutions using FDTD. Resolutions are quoted on each sub-plot in units of pixels-per-wavelength (ppw).}
\label{validationPlot}
\end{figure}

The accuracy of the FDTD simulations was estimated by using them to calculate the absolute value of the coherence $\rho_{12}$ in vacuum, which is known to be identically zero (see section \ref{VacuumSection} and Ref.~\cite{Agarwal2000}). The deviation from zero can then be used to estimate the errors introduced by the numerical nature of the method. A data set was generated by randomly sampling points from within the simulation box and calculating $|\rho_{12}|$ at each. As shown in Fig.~\ref{resolutionPlot}, these displayed a systematic resolution-dependent displacement from zero, as well as random fluctuations around that value. The mean value was therefore used as a systematic error, while the standard deviation was taken as a random error, which were subsequently combined in quadrature to give an overall error. Enough simulations were run so that the total error reached a steady value, as demonstrated in Fig.~\ref{resolutionPlot}.
To test the validity of these error bounds we simulated the case of the perfect reflector and compared with the analytic result in Eq.~\eqref{PMResult}. This is shown in Fig.~\ref{validationPlot}, where the sizes of the error bars are correspond to each resolution shown on Fig.~\ref{resolutionPlot}.  
From this it was determined that the resolution giving the best tradeoff between computational overhead and accuracy was 12 pixels per wavelength. This was used for the simulations in the main text, in which all errors are less than or similar to the thickness of the lines on the plots.


%

\end{document}